# Crystal Field-Phonon Coupling in $ErFeO_3$ Single Crystal Via Inelastic Neutron Scattering

*Guochu Deng,[1*†] Xiaoxuan Ma,[2†] Wanting Yang,[2†] Wei Ren,[2] Shixun Cao,[2*] Zhenxiang Cheng,[3] Garry J. McIntyre[1]*

[1]*Australian Centre for Neutron Scattering, Australian Nuclear Science and Technology Organisation, New Illawarra Road, Lucas Heights NSW 2234, Australia*

[2] *Materials Genome Institute, Shanghai University, Shanghai 200444, China*

[3] *Australian Institute for Innovative Materials, Institute for Superconducting and Electronic Materials, University of Wollongong, Squires Way, North Wollongong, NSW, 2500, Australia*

**Abstract**

Crystal-field (CF) excitations and their coupling to lattice vibrations were investigated in $ErFeO_3$ using inelastic neutron scattering. Pronounced modifications of the excitation spectra, including energy shifts, intensity redistribution, and linewidth changes, were observed as a consequence of CF-phonon coupling. Significant renormalization of the first and second CF excitations, located near 5 meV and 12 meV, respectively, was detected along both transverse and longitudinal directions in the vicinity of the reciprocal-space point $Q$ = (2, 0, 4). The coupling effects become more prominent as the temperature increases from 300 K to 650 K, while a slight suppression is observed at 750 K. Remarkably, CF excitation peaks that are strongly weakened by thermal population effects re-emerge in specific momentum regions through interaction with phonon modes. The coupled excitations exhibit characteristic signatures of hybridization, including energy renormalization, intensity enhancement, and linewidth narrowing. Density-functional-theory-based phonon calculations further support the interpretation of CF-phonon coupling in $ErFeO_3$. These results demonstrate that strong interactions between CF excitations and lattice vibrations can substantially modify the excitation spectra in rare-earth orthoferrites.

* Corresponding authors' emails: guochu.deng@ansto.gov.au, sxcao@shu.edu.cn;
† These authors contributed equally.

## Introduction

In crystalline solids, the 3d orbitals of transition-metal ions and the 4*f* orbitals of rare-earth ions are influenced by the electric fields generated by surrounding ligands. This interaction, commonly referred to as the crystal-field (CF) effect, lifts the orbital degeneracy and splits the electronic states into a series of discrete energy levels. Owing to the shielding provided by the outer 5*s* and 5*p* electrons, the 4*f* electrons of rare-earth ions experience a substantially weaker crystal field than the 3*d* electrons of transition-metal ions. Consequently, CF excitation energies in rare-earth compounds typically range from a few meV to several hundred meV, overlapping significantly with the energy scale of lattice vibrations. Because both CF excitations and phonons are influenced by the local ligand environment, strong interactions between these two degrees of freedom can arise. CF-phonon coupling is therefore expected to play an important role in many rare-earth materials.

One of the earliest experimental demonstrations of CF-phonon coupling was reported by M. Loewenhaupt and co-workers[1] in $CeAl_2$ using inelastic neutron scattering. Two anomalous excitation peaks were observed that could not be explained within a conventional CF model. Subsequent theoretical work attributed the splitting of the CF excitation to the formation of vibronic bound states arising from strong magnetoelastic coupling between CF excitations and lattice vibrations.[2] Theoretical analyses further showed that the emergence of hybrid CF-phonon states depends strongly on the magnitude of the magnetoelastic coupling constant, with sufficiently strong coupling leading to pronounced level repulsion and excitation splitting. The unusually large magnetoelastic coupling in $CeAl_2$ explains why such splitting is not commonly observed in other members of the $REAl_2$ family.

Since the pioneering work on $CeAl_2$, CF-phonon coupling has been reported in a wide range of materials. In $CeCu_2$, anomalous temperature-dependent excitation softening observed by inelastic neutron scattering was attributed to interactions between CF and phonon modes.[3] Similar behaviour has been reported in $YbPO_4$ through both Raman-scattering and neutron-scattering measurements.[4, 5] Evidence for CF-phonon coupling has also been identified in the high-temperature superconductor $NdBa_2Cu_3O_{7-\delta}$, where Raman measurements revealed an unexpected double-peak feature associated with coupled CF and phonon excitations.[6, 7] More recently, hybrid CF-phonon excitations and vibronic bound states have been reported in spin-ice systems including $Ho_2Ti_2O_7$,[8] $Ho_2Sn_2O_7$[9] and $Ho_2Ge_2O_7$[10], as well as in $Ce_2O_3$,[11] $CeAuAl_3$,[12] and $TbInO_3$.[13] Collectively, these studies demonstrate that CF-phonon interactions are a widespread phenomenon in rare-earth compounds and can give rise to rich hybrid excitation behaviour.

The observation of CF-phonon coupling is particularly favourable in rare-earth systems because CF excitation energies often coincide with low-energy acoustic phonon branches. As a result, hybridization between electronic and lattice excitations can occur over a broad region of reciprocal space. Despite numerous reports of CF-phonon coupling in rare-earth materials, detailed studies of its temperature dependence and momentum-space evolution remain relatively limited.

$ErFeO_3$ is a member of the rare-earth orthoferrite family $REFeO_3$. The $Fe^{3+}$ magnetic sublattice develops a canted antiferromagnetic order below approximately 670 K, whereas long-range antimagnetic ordering of the $Er^{3+}$ moments occurs near 4.2 K. In addition, a spin-reorientation transition takes place between approximately 88 K and 100 K. The $Er^{3+}$ ion possesses a $^7I_{15/2}$ ground-state multiplet that is split into a series of CF doublets. In our previous work,[14] the CF scheme of $ErFeO_3$ was determined from inelastic neutron scattering measurements and CF modelling. The first and second CF excitations were identified near 5.5 meV and 12.5 meV, respectively, placing them within the energy range of low-lying acoustic phonons and making $ErFeO_3$ an attractive candidate for investigating CF-phonon interactions.

In this work, we investigate CF-phonon coupling in $ErFeO_3$ by means of inelastic neutron scattering measurements performed at three temperatures. The results reveal pronounced interactions between CF excitations and both transverse and longitudinal acoustic phonons. These interactions lead to substantial modifications of excitation energies, spectral intensities, and linewidths within specific regions of momentum space. The coupling effects become increasingly evident when the phonon and CF excitation energies approach each other, resulting in characteristic signatures of mode hybridization. Our findings provide evidence for strong CF-phonon coupling in $ErFeO_3$ and highlight the broader importance of such interactions in rare-earth compounds.

## II. Experiment and Calculation

### Experimental Details

A high-quality single crystal of $ErFeO_3$ was grown using an optical travelling-solvent floating-zone furnace at Shanghai University. The crystal quality and orientation were verified using the neutron

Laue diffractometer Joey at the Australian Centre for Neutron Scattering (ACNS), ANSTO. Inelastic neutron scattering (INS) measurements were performed on the thermal-neutron triple-axis spectrometer Taipan operating in constant-$E_f$ mode with an open-40′-40′-open collimation configuration.[15] A closed-cycle cryofurnace was employed for temperature control. The experimental spectra were fitted using resolution-convoluted models implemented in the Python packages TasVisAn and InsPy.[16]

***DFT Calculation***

To investigate the phonon dispersions of $ErFeO_3$, density functional theory (DFT) calculations were performed within the projector augmented-wave (PAW) formalism using the Vienna Ab Initio Simulation Package (VASP). The Hubbard-corrected generalized gradient approximation (GGA+U) [17-19]with the Perdew-Burke-Ernzerhof exchange-correlation functional for solids (PBEsol) was employed.[20-22] The on-site Coulomb interaction parameter U was set to 5.0 eV for Er and 4.5 eV for Fe. A plane-wave cutoff energy of 600 eV and a Monkhorst-Pack 5×5×3 k-point mesh were used. Phonon dispersions were calculated using the finite-displacement method implemented in Phonopy,[23, 24]and the resulting phonon spectra were visualized using Euphonic.[25]

## III. Results and Discussions

### A) CF and transversion acoustic (TA) phonon coupling in the *Q*(2, 0, 4) zone

Figure 1 shows INS spectra measured at a series of transverse positions (2+$q$, 0, 4-$q$) ($q$ = 0.1 ~ 0.5) near the zone centre $Q$ = (2, 0, 4) at 300, 650, and 750 K. The fitted excitation energies, intensities, and linewidths are summarized in Fig. 2. The most intense mode exhibits a clear dispersion and rapidly decreasing intensity with increasing $q$, consistent with a transverse acoustic (TA) phonon branch.

At 300 K, both CF and phonon excitations are observed. The strongest peak shows a clear dispersion with the increase of $q$, which can be safely assigned to the TA phonon. At the positions with $q > 0.1$, another two peaks appear at 5.5 meV and 12.5 meV, showing less variation with momentum transfer, which correspond to the first and second CF excitations, respectively, consistent with our previous CF report. Fitting the peaks in the scans carefully, the difference between the first and second CF excitation peaks shows up. While the first CF excitation shows little variation with momentum transfer, the second CF excitation undergoes a noticeable softening for $q \geq 0.3$, shifting from approximately 12.5 meV to 11.8 meV at $q = 0.5$. Simultaneously, its linewidth becomes narrower. These observations suggest an interaction between the second CF excitation and the TA phonon branch in this momentum range.

At 650 K and 750 K, thermal population effects strongly suppress the intrinsic CF transitions, while the TA phonon intensity increases substantially. The main thermal impact causes the disappearance of the first CF excitation peak at 5.5 meV at these two temperatures. For the second CF excitation, the peak is hardly to be seen at the low $q$ range (q = 0.1 and 0.2). Surprisingly, despite the expected weakening of the second CF excitation, a well-defined excitation remains visible near 11.8 meV for $q = 0.3 \sim 0.5$ at 650 K and for $q = 0.4$ and 0.5 at 750 K. Remarkably, this peak becomes sharper and more intense than that observed at 300 K. The enhanced peak intensity and narrowed linewidth at these higher $q$ positions demonstrate high contrast to the corresponding strongly broadened or nearly absent feature at lower $q$ values. Such momentum-dependent behaviour cannot be explained by thermal effects alone and is consistent with enhanced CF-phonon coupling when the phonon and CF excitation energies become comparable.

Particular attention should be paid to the second CF excitation in the $q > 0.3$ region at 650 and 750 K. In this $q$ range, the CF excitation develops a weak but distinct dispersion that follows the trend of the TA phonon dispersion (Fig. 2(a)). In particular, the CF peak centre at $q = 0.5$ slightly higher in energy than that at $q = 0.4$, giving the CF excitation a positive dispersion similar to that of the TA phonon. This behaviour is markedly different from the nearly $q$-independent CF excitation expected for a localized CF transition and provides evidence that the CF excitation is strongly coupled to the phonon mode.

Such behaviour can be understood in terms of CF–phonon hybridization. When a CF excitation and a phonon approach each other in energy and are coupled, the eigenstates of the system are no longer pure CF or phonon excitations. Instead, they become hybrid modes that can acquire spectral characteristics from both unperturbed excitations. Within perturbation theory, this coupling can produce several observable effects, including energy renormalization, level repulsion (avoided crossing), intensity redistribution or intensity borrowing, and changes in the excitation linewidth.[2] Consequently, a hybrid mode can inherit part of the dispersion and dynamical character of the phonon, even though the corresponding uncoupled CF transition would be essentially dispersionless.

The experimental observations are consistent with this picture. In the $q > 0.3$ region, the second CF excitation acquires a dispersion with a slope comparable to that of the TA phonon, indicating that the CF excitation has acquired phonon-like character. At the same time, the CF excitation exhibits enhanced spectral intensity and a pronounced reduction in linewidth. The enhanced intensity can be understood as intensity borrowing from the strongly scattering phonon mode, while the linewidth change reflects the modified lifetime of the hybrid excitation. Together with the correlated evolution of the CF and TA-phonon branches, these observations provide strong evidence for substantial CF–phonon hybridization in this *q* range.

If the observed branches exhibit a clear repulsion rather than simply following parallel dispersions, this behaviour can be further identified as an avoided crossing. In such a case, the two hybridized branches bend away from one another near the nominal crossing point, producing a finite anti-crossing gap. The observed branch repulsion, shared dispersion characteristics, intensity redistribution, and linewidth modification would then provide complementary signatures of the same underlying CF–phonon coupling. Thus, the second CF excitation and the TA phonon undergo strong hybridization in this specific *q* region, resulting in pronounced modifications of both the CF and phonon excitations.

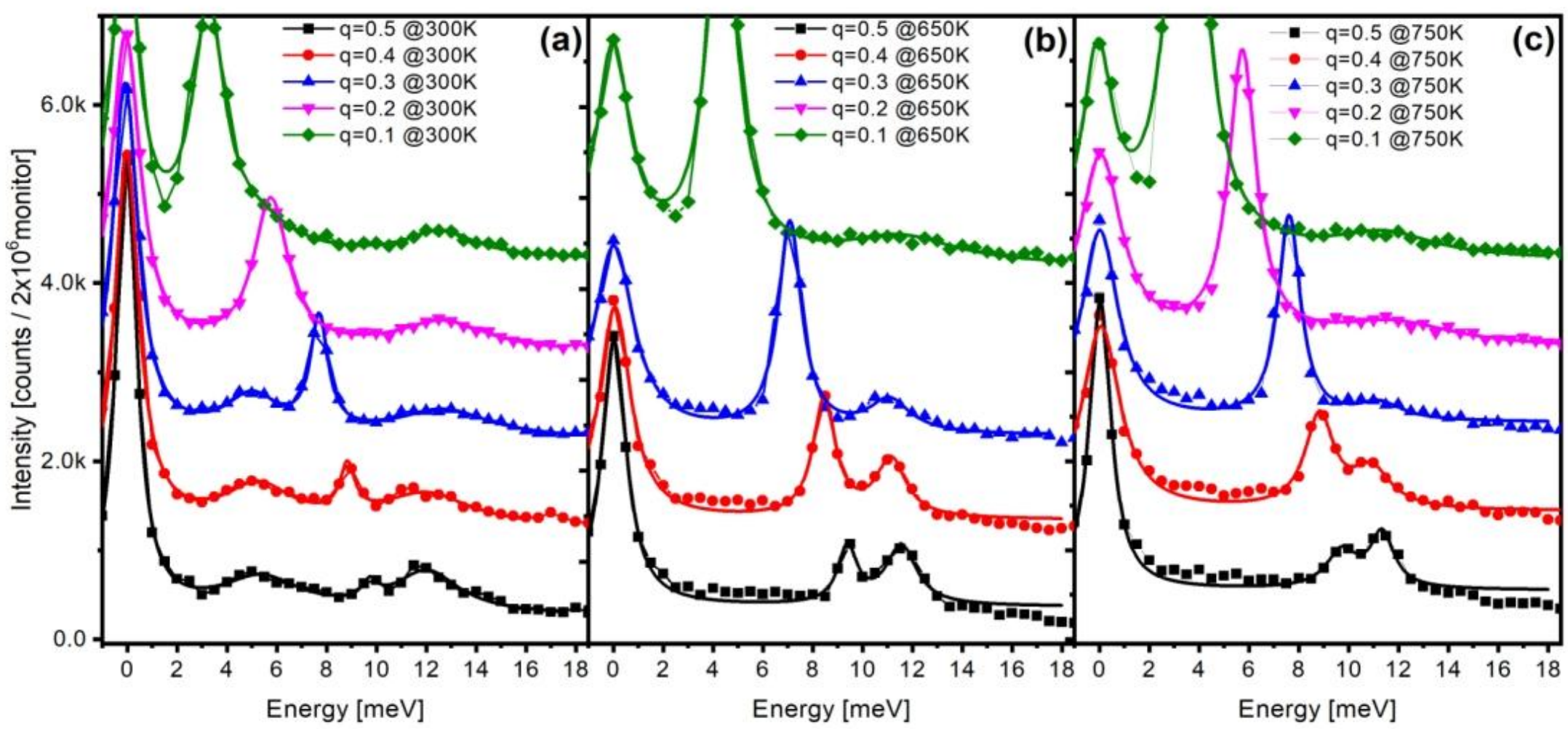

Fig. 1 The energy scans at a series of transverse *q* positions (2+q, 0, 4-q) measured at (a) 300K, (b) 650k, and (c) 750 K.

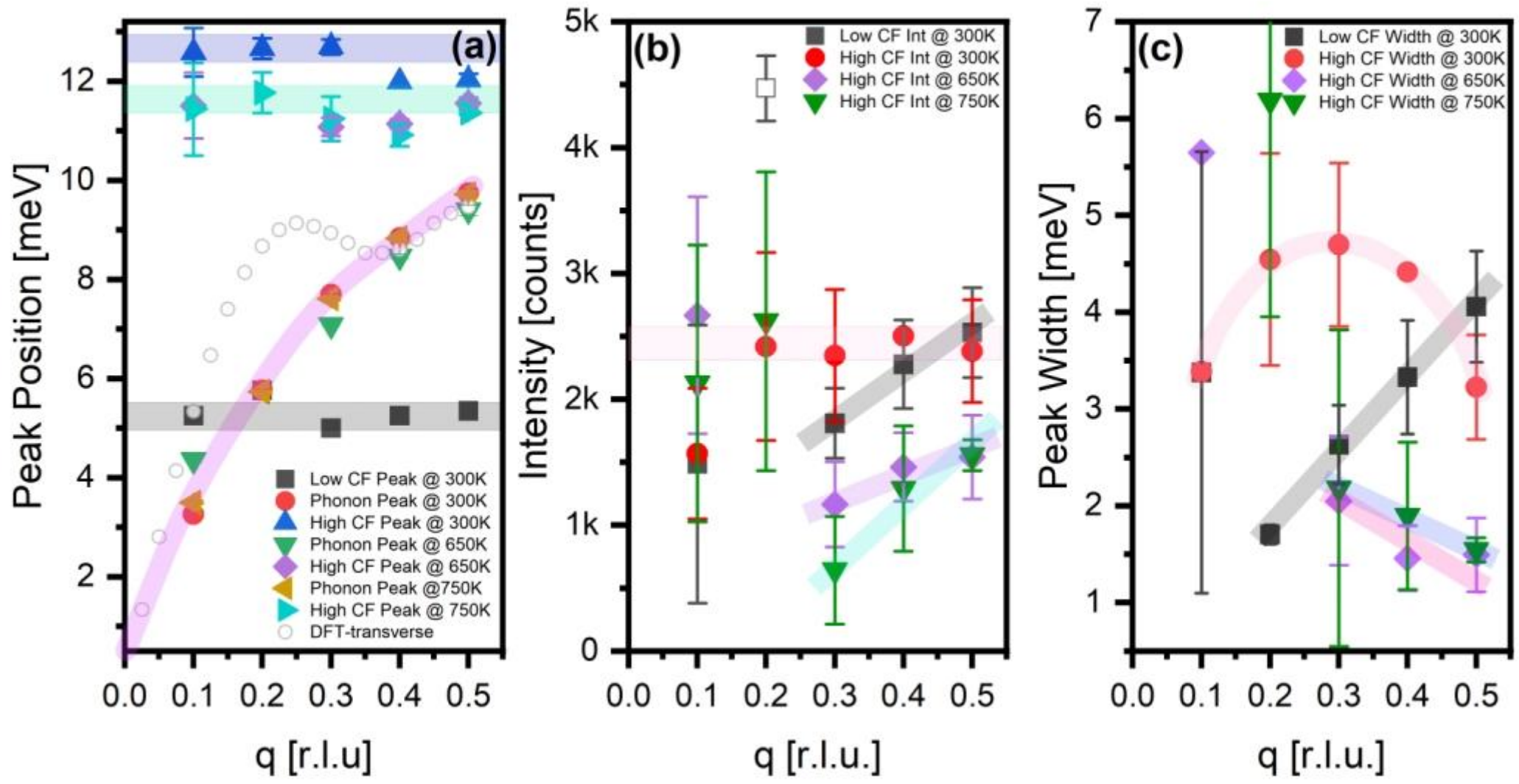

Fig. 2 (a) Excitation energies, (b) peak intensities, and (c) peak widths of the energy scans at different transverse $q$ positions at 300 K, 650 K, and 750 K obtained from the experimental data fitting in Fig. 1. The theoretical phonon dispersion along this direction were calculated from the DFT calculation and are shown in (a).

### B) CF-longitudinal acoustic (LA) phonon coupling

Energy scans were also performed at a series of longitudinal positions $Q$ (2-$q$, 0, 4-2$q$) with $q$ = 0.1 ~ 0.5 at 300, 650, and 750 K. The experimental data and corresponding resolution-convoluted fits are shown in Fig. 3, while the extracted excitation energies, intensities, and linewidths are summarized in Fig. 4. Similar to the transverse measurements, multiple excitation peaks are observed and assigned to CF and phonon modes. Owing to the rapid decrease in phonon intensity with increasing momentum transfer, the longitudinal acoustic (LA) phonon can be clearly identified only in the $q$ = 0.1 scan.

The 300 K measurements are shown in Fig. 3(a). Two CF excitation peaks are observed at approximately 5.3 meV and 12.5 meV throughout the measured $q$ range. At $q$ = 0.1, an LA phonon is detected near 3.7 meV and partially overlaps with the first CF excitation. As $q$ increases from 0.2 to 0.5, both CF excitation energies remain nearly unchanged, and no obvious anomalies are observed in either their intensities or linewidths. These results suggest that the coupling between the CF excitations and the LA phonon is relatively weak at 300 K.

The excitation spectra change significantly at 650 K and 750 K. At $q$ = 0.1, the second CF excitation expected near 12.5 meV is no longer observed, whereas the LA phonon near 3.7 meV becomes substantially more intense. At $q = 0.2$, two excitation peaks emerge at both temperatures. The lower-energy peak remains close to 5.3 meV and can be assigned to the first CF excitation. The higher-energy peak appears near 10 meV, considerably below the position of the second CF excitation observed in the transverse scans. Comparison with the calculated longitudinal phonon dispersion indicates that this feature is consistent with the LA phonon mode. Its enhanced visibility at elevated temperatures is likely associated with the increased phonon population.

A particularly interesting evolution occurs at $q = 0.3$. At both 650 K and 750 K, the first CF excitation shifts upward from approximately 5.3 meV to about 7 meV. At the same time, the higher-energy excitation moves to around 12 meV. Based on its energy and evolution, this high energy feature is attributed to the second CF excitation because the intensity contribution of the LA phonon at this $q$ position is expected to be very weak according to the DFT calculations.

At $q = 0.4$ and 0.5, the evolution of the second CF excitation becomes even more pronounced. At 750 K, the peak softens to approximately 11.4 meV, while the first CF excitation becomes strongly suppressed and eventually disappears. Although the LA phonon is not directly resolved in this momentum range because of its weak intensity, the calculated dispersion indicates that its energy decreases towards the zone boundary. The observed CF excitation follows a similar trend, displaying a dispersion that closely resembles the calculated LA branch. The persistence of the CF peak at high $q$, together with its phonon-like momentum dependence, provides strong evidence that the second CF excitation remains coupled to the LA phonon even when the phonon itself is no longer directly observable.

The behaviour of the first CF excitation also supports the CF-LA coupling scenario. At $q = 0.2$, the excitation remains visible near 5.3 meV at both 650 K and 750 K. However, at $q = 0.3$, the peak shifts to approximately 6.9 meV and exhibits a reduction in intensity. The excitation can still be identified at $q = 0.4$ for 650 K but becomes nearly invisible at 750 K, and it is absent at $q = 0.5$ for both temperatures. This evolution suggests that the excitation does not behave as an isolated CF transition. Instead, its energy and spectral weight vary systematically with $q$, consistent with interactions with the LA phonon branch. A plausible interpretation is that the first CF excitation merges with the low-energy LA phonon near $q = 0.1$, gradually separates at intermediate $q$, and becomes progressively weakened as the LA phonon intensity decreases towards the zone boundary. In contrast, the nearly momentum-independent intensity observed at 300 K is characteristic of an uncoupled CF excitation. The $q$-dependent spectral redistribution observed at elevated temperatures therefore provides further support for CF-LA phonon hybridization.

Additional support for the coupling between the second CF excitation and the LA phonon is provided by the persistence of the high-energy CF peak near the zone boundary. According to the DFT calculations shown in Fig. 5(b), the LA phonon intensity decreases substantially with increasing $q$ and becomes extremely weak near the zone boundary. The excitation observed near 11 meV is therefore most naturally assigned to the second CF transition. Notably, its energy remains relatively close to that predicted for the LA phonon branch, suggesting that its spectral properties continue to be influenced by CF-phonon interactions. The combination of energy renormalization and sustained spectral intensity is consistent with a hybridized excitation picture.

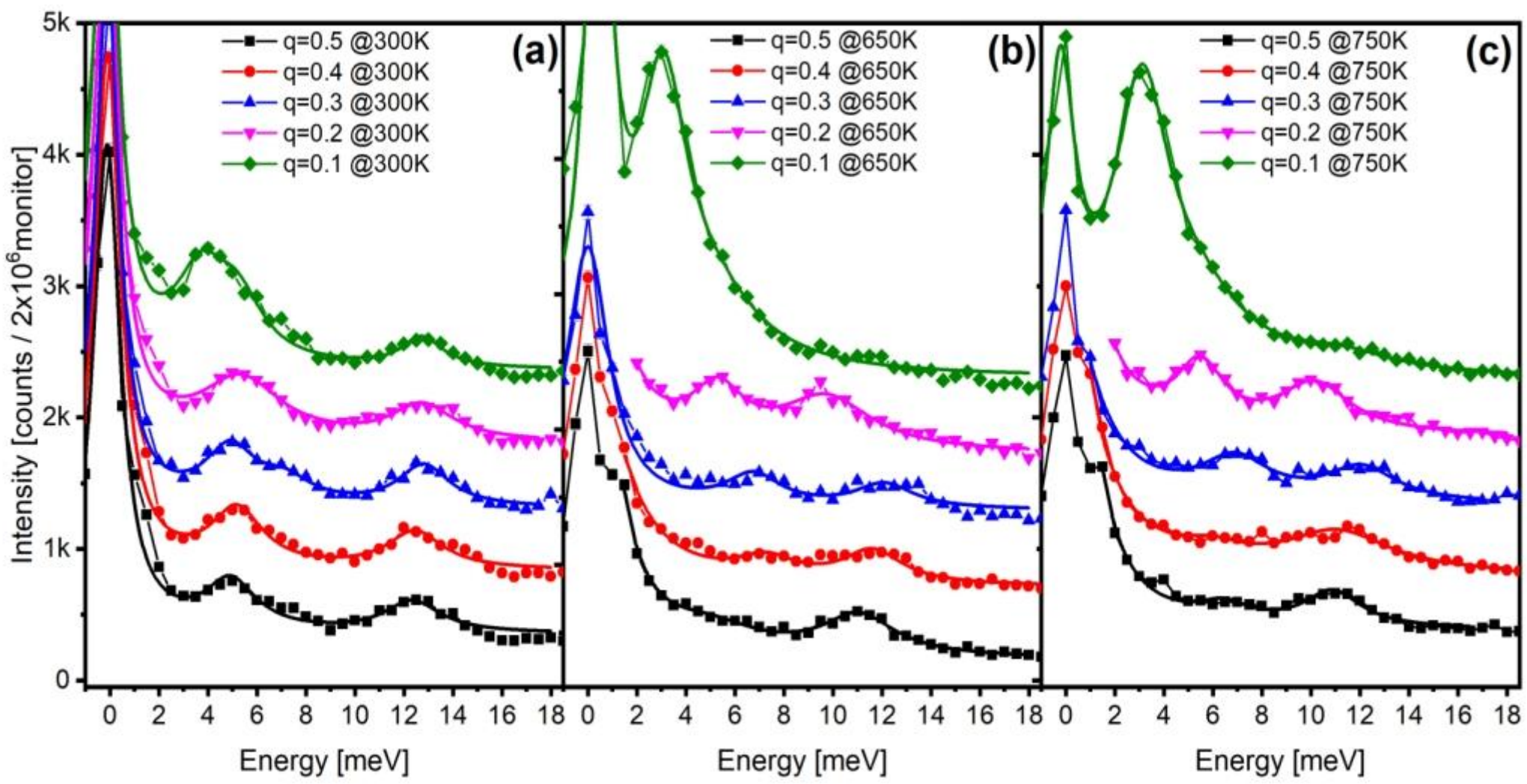

Fig. 3 The energy scans at a series of longitudinal $q$ positions (2-$q$, 0, 4-2$q$) measured at (a) 300 K, (b) 650 K, and (c) 750 K.

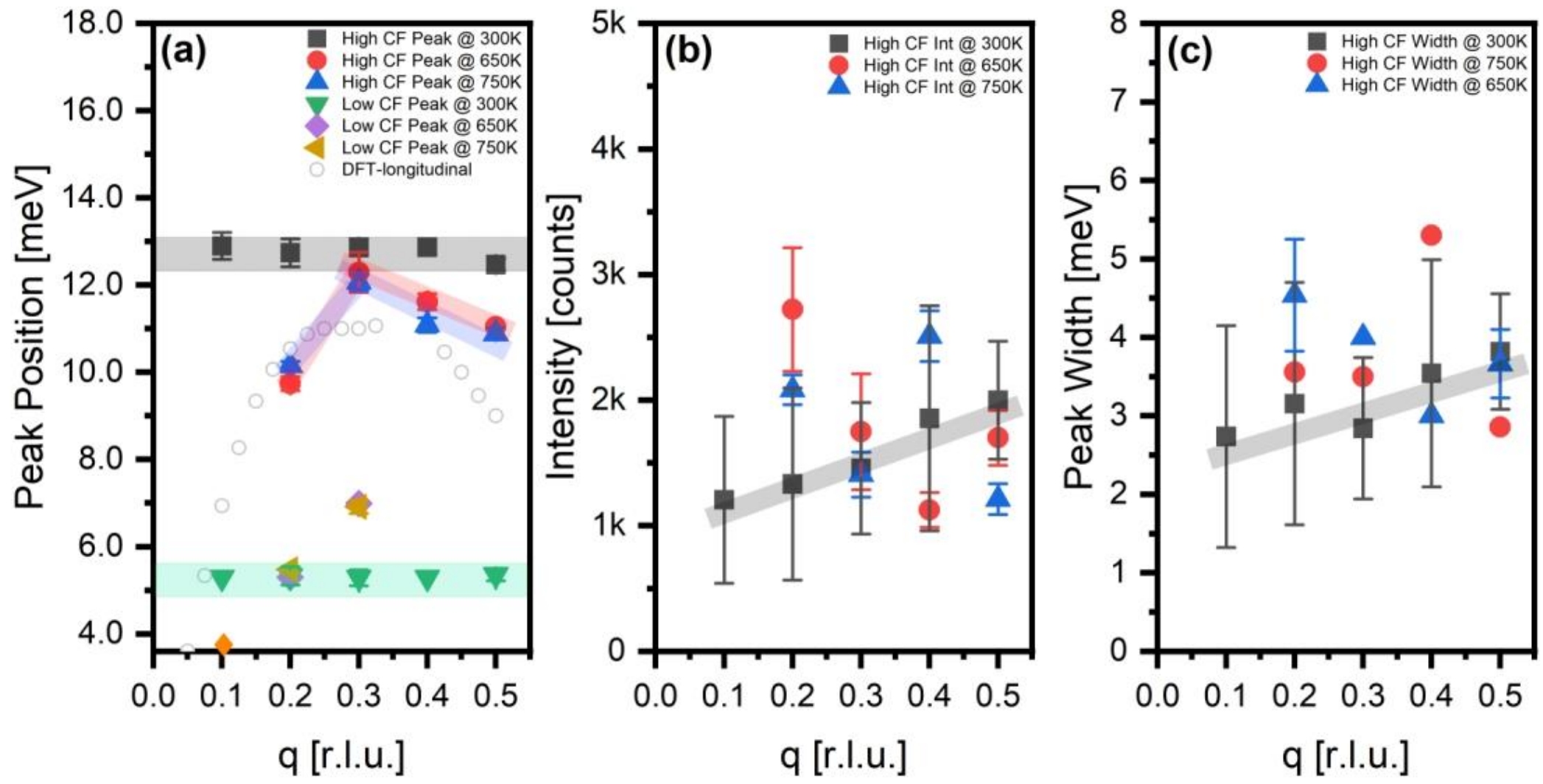

Fig. 4 (a) Excitation energies, (b) peak intensities, and (c) peak widths of the energy scans at different longitudinal $q$ positions at 300 K, 650 K, and 750 K obtained from the experimental data fitting in Fig. 3. The theoretical phonon dispersion along this direction were calculated from the DFT calculation and are shown in (a).

### *C. Overall Characteristics of the CF-Phonon Coupling*

The observations obtained along both the transverse and longitudinal directions reveal a consistent picture of CF-phonon hybridization in $ErFeO_3$. At elevated temperatures, the CF excitations exhibit pronounced energy renormalization, anomalous intensity enhancement, and linewidth narrowing when their energies approach those of the acoustic phonons. In several momentum regions, the CF excitations also develop phonon-like dispersion, indicating that the two modes are no longer independent but become hybridized through magnetoelastic interactions. These effects are particularly striking because the intrinsic thermal population of excited CF levels is expected to suppress the neutron-scattering intensity of CF transitions. Instead, well-defined CF-related excitations persist and, in some regions, become more prominent at high temperatures. Such behaviour strongly suggests a transfer of spectral weight from the phonon branch to the CF excitations through mode hybridization. The experimental observations, together with the DFT-calculated phonon dispersions, consistently support a CF-phonon coupling scenario in $ErFeO_3$.

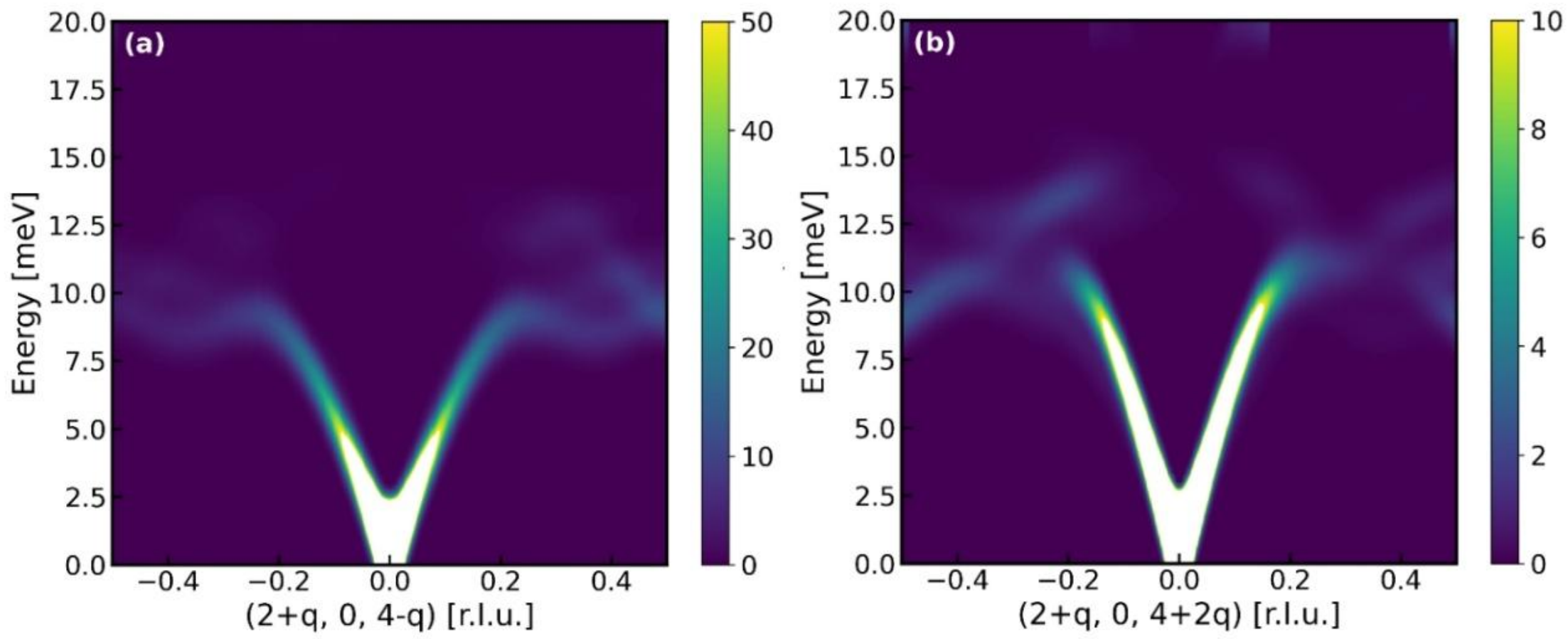

Fig. 5 Theoretical (a) transverse and (b) longitudinal phonon dispersions calculated through the DFT calculation.

## IV. Conclusion

We have investigated the interaction between CF excitations and lattice vibrations in an $ErFeO_3$ single crystal using inelastic neutron scattering and DFT phonon calculations. Pronounced coupling effects were observed near the reciprocal-space point $Q$ = (2, 0, 4), where the energies of the CF excitations become comparable to those of the acoustic phonons. The coupling manifests itself through energy renormalization, linewidth narrowing, intensity redistribution, and phonon-like dispersion of the CF excitations. These signatures were observed for both transverse and longitudinal acoustic phonon branches and become considerably more pronounced at elevated temperatures.

A particularly notable result is the persistence and enhancement of CF-related spectral features in momentum regions where thermal population effects would otherwise strongly suppress the CF transitions. This behaviour is consistent with hybridization between CF and phonon excitations, which enables substantial spectral-weight transfer between the two modes. The overall experimental and theoretical results therefore provide compelling evidence for strong CF-phonon coupling in $ErFeO_3$.

More broadly, the present work demonstrates that low-energy CF excitations in rare-earth orthoferrites can couple strongly to acoustic phonons over an extended region of reciprocal space. These findings highlight the important role of CF-phonon interactions in shaping the excitation spectra of rare-earth materials and establish $ErFeO_3$ as a promising model system for studying hybrid electronic-lattice excitations.

**Acknowledgement**

The authors would like to express their sincere gratitude to Dr. Kirrily Rule at Australian Centre for Neutron Scattering, ANSTO for her thoughtful discussions and insightful comments, which have contributed to improving this manuscript. This work is partially supported by the National Natural Science Foundation of China (NSFC, No. 12374116, 12674145), and the China Postdoctoral Science Foundation (No. 2024M751931, 2026M793718). The authors would like to thank the Australian Centre for Neutron Scattering, ANSTO for the allocation of beam time on Taipan (P3169).

**References**

[1] M. Loewenhaupt, B.D. Rainford, F. Steglich, Dynamic Jahn-Teller Effect in a Rare-Earth Compound: CeAl2, Physical Review Letters, 42 (1979) 1709–1712.

[2] P. Thalmeier, P. Fulde, Bound State between a Crystal-Field Excitation and a Phonon in CeAl2, Physical Review Letters, 49 (1982) 1588–1591.

[3] U. Witte, S. Kramp, M. Braden, P. Svoboda, M. Loewenhaupt, Crystal field–phonon coupling in the Kondo lattice CeCu2, Applied Physics A, 74 (2002) s571–s573.

[4] P.C. Becker, G.M. Williams, N.M. Edelstein, J.A. Koningstein, L.A. Boatner, M.M. Abraham, Observation of strong electron-phonon coupling effects in YbPO4, Physical Review B, 45 (1992) 5027–5030.

[5] C.K. Loong, M. Loewenhaupt, J.C. Nipko, M. Braden, L.A. Boatner, Dynamic coupling of crystal-field and phonon states in YbPO4, Physical Review B, 60 (1999) R12549–R12552.

[6] E.T. Heyen, R. Wegerer, M. Cardona, Coupling of phonons to crystal-field excitations in NdBa2Cu3O7-d, Physical Review Letters, 67 (1991) 144–147.

[7] T. Ruf, E.T. Heyen, M. Cardona, J. Mesot, A. Furrer, Raman study of crystal-field--phonon coupling in NdBa2Cu3O7-d in a magnetic field, Physical Review B, 46 (1992) 11792–11797.

[8] J. Gaudet, A.M. Hallas, C.R.C. Buhariwalla, G. Sala, M.B. Stone, M. Tachibana, K. Baroudi, R.J. Cava, B.D. Gaulin, Magnetoelastically induced vibronic bound state in the spin-ice pyrochlore Ho2Ti2O7, Physical Review B, 98 (2018) 014419.

[9] K. Matsuhira, Y. Hinatsu, K. Tenya, T. Sakakibara, Low temperature magnetic properties of frustrated

pyrochlore ferromagnets Ho2Sn2O7 and Ho2Ti2O7
J. Phys.: Condens. Matter, 12 (2000) L649
[10] H.D. Zhou, J.G. Cheng, A.M. Hallas, C.R. Wiebe, G. Li, L. Balicas, J.S. Zhou, J.B. Goodenough, J.S. Gardner, E. S. Choi, Chemical Pressure Effects on Pyrochlore Spin Ice, Phys. Rev. Lett., 108 (2012) 207206
[11] A. Sethi, J.E. Slimak, T. Kolodiazhnyi, S.L. Cooper, Emergent Vibronic Excitations in the Magnetodielectric Regime of Ce2O3, Physical Review Letters, 122 (2019) 177601.
[12] P. Čermák, A. Schneidewind, B. Liu, M.M. Koza, C. Franz, R. Schönmann, O. Sobolev, C. Pfleiderer, Magnetoelastic hybrid excitations in CeAuAl3, Proceedings of the National Academy of Sciences, 116 (2019) 6695–6700.
[13] M. Ye, X. Xu, X. Wang, J. Kim, S.-W. Cheong, G. Blumberg, Crystal-field excitations and vibronic modes in the triangular-lattice spin-liquid candidate TbInO3, Physical Review B, 104 (2021) 085102.
[14] G. Deng, Internal and external field effects upon crystal field excitations in REFeO3 (RE = Nd3+, Er3+, Yb3+, Pr3+, and Ho3+), Physica B: Condensed Matter, 726 (2026) 418310.
[15] S.A. Danilkin, M. Yethiraj, TAIPAN: Thermal Triple-Axis Spectrometer, Neutron News, 20 (2009) 37–39.
[16] G. Deng, G.J. McIntyre, TasVisAn and InsPy: Python packages for triple-axis spectrometer data visualization, analysis, instrument resolution calculation and convolution, Journal of Applied Crystallography, 58 (2025) 2149–2162.
[17] G. Kresse, J. Furthmüller, Efficient iterative schemes for ab initio total-energy calculations using a plane-wave basis set, Physical Review B, 54 (1996) 11169–11186.
[18] G. Kresse, D. Joubert, From ultrasoft pseudopotentials to the projector augmented-wave method, Physical Review B, 59 (1999) 1758–1775.
[19] S.L. Dudarev, G.A. Botton, S.Y. Savrasov, C.J. Humphreys, A.P. Sutton, Electron-energy-loss spectra and the structural stability of nickel oxide: An LSDA+U study, Physical Review B, 57 (1998) 1505–1509.
[20] J.P. Perdew, Y. Wang, Accurate and simple analytic representation of the electron-gas correlation energy, Physical Review B, 45 (1992) 13244–13249.
[21] J.P. Perdew, K. Burke, M. Ernzerhof, Generalized Gradient Approximation Made Simple, Physical Review Letters, 77 (1996) 3865–3868.
[22] J.P. Perdew, A. Ruzsinszky, G.I. Csonka, O.A. Vydrov, G.E. Scuseria, L.A. Constantin, X. Zhou, K. Burke, Restoring the Density-Gradient Expansion for Exchange in Solids and Surfaces, Physical Review Letters, 100

(2008) 136406.

[23] A. Togo, L. Chaput, T. Tadano, I. Tanaka, Implementation strategies in phonopy and phono3py, Journal of Physics: Condensed Matter, 35 (2023) 353001.

[24] A. Togo, First-principles Phonon Calculations with Phonopy and Phono3py, Journal of the Physical Society of Japan, 92 (2023) 012001.

[25] R. Fair, A. Jackson, D. Voneshen, D. Jochym, D. Le, K. Refson, T. Perring, Euphonic: inelastic neutron scattering simulations from force constants and visualization tools for phonon properties, Journal of Applied Crystallography, 55 (2022) 1689–1703.